# Highly active hydrogen evolution facilitated by topological surface states on a Pd/SnTe metal/topological crystalline insulator heterostructure

Qing Qu[1,2,†], Bin Liu[1,†], Wing Sum Lau[1], Ding Pan[1,3,4,*] and Iam Keong Sou[1,2,*]

[1]Department of Physics, The Hong Kong University of Science and Technology, Clear Water Bay, Hong Kong, China.

[2]William Mong Institute of Nano Science and Technology, The Hong Kong University of Science and Technology, Hong Kong, China.

[3]Department of Chemistry, The Hong Kong University of Science and Technology, Hong Kong, China.

[4] HKUST Shenzhen-Hong Kong Collaborative Innovation Research Institute, Shenzhen, China.

[†] These authors contributed equally to this work.

[*] Corresponding authors

# Abstract

Recently, topological quantum materials have emerged as a promising electrocatalyst for hydrogen evolution reaction (HER). However, most of their performance largely lags behind noble metals such as benchmark platinum (Pt). In this work, a Pd(20nm)/SnTe(70nm) heterostructure, fabricated by molecular beam epitaxy and electron beam evaporation, is found to display much higher electrocatalytic activity than that of a pure Pd(20nm) thin film and even higher than that of a commercial Pt foil. This heterostructure adopts an extracted turnover frequency value more than two times higher than that of the Pd(20nm) thin film at a potential of 0.2 V, indicating a much higher intrinsic activity per Pd site. Density functional theory calculations show that the conventional d-band theory, which works well for many transition metal heterostructures, cannot explain the enhancement of electrocatalytic performance. Instead, we found that the topological surface states (TSSs) of the SnTe (001) underlayer play a key role; electrons transfer from both the Pd surface and the adsorbed H atoms to the TSSs of SnTe (001), resulting in weaker Pd-H binding strength and more favorable hydrogen adsorption free energies. Our work demonstrates for the first time that a metal/topological quantum material heterostructure could be a prominent catalyst to enjoy HER activity outperforming that of a commercial Pt foil and offers a promising direction to optimize the performance of electrocatalysts based on topological quantum materials.

# Introduction

The most essential prerequisite for the hydrogen economy is sustainable hydrogen production. The hydrogen evolution reaction ($2H^+ + 2e^- \rightarrow H_2$, HER) by electrocatalytic water splitting that occurs at the solid-liquid interface plays an essential role in realizing clean energy conversion. Platinum (Pt) is still considered as the benchmark catalyst for HER with its excellent activity.[1] However, in addition to the scarcity, Pt has poor electrochemical stability, which is associated with leaching in corrosive electrolytes and irreversible aggregation of Pt nanoparticles[2,3], limiting its practical applications. According to the volcano plot[4], Palladium (Pd) can be used as a substitute for Pt with relatively good HER activity. Regarding thermal deposition of Pd, a standard growth rate of 1 Å/s only requires a source temperature of 1,200 $^0$C while the corresponding source temperature for Pt would be about 1,750 $^0$C, the latter usually presents a more challenging issue for physical vapor deposition techniques. Thus in this work, Pd has been used as a metal catalyst layer in our heterostructure samples.

In searching for highly efficient HER electrode materials, topological insulators (TIs) as a new class of quantum matter have recently attracted increasing attention. In particular, our group reported that $Bi_2Te_3$ TI thin films with partially oxidized surfaces or Te vacancies can have high HER activities, which is attributed to the electrons in the H atoms being transferred to the metallic surface states, so the H atoms become easier to adhere to the $Bi_2Te_3$ surfaces.[5] Our group also found that topological crystalline insulators (TCIs) SnTe (001) and SnTe (111) enjoy comparable good intrinsic activity while that of SnTe (211) with absent or fragile TSSs is significantly lower, which is attributed to the enhanced charge transfer between H atoms and TSSs.[6] Based on first-principles density functional theory (DFT) calculations, Chen et al.[7] proposed a novel application of 3D TIs in heterogeneous catalysis. Their calculations show that the TSSs effectively act as an electron bath, making oxygen and carbon monoxide molecules more prone to dissociate and get adsorbed on a gold-covered $Bi_2Se_3$ structure through enhancing the adsorption energies of these molecules. Then our group presented results of experimental studies on a similar structure, Pd-covered $Bi_2Te_3$, focusing on the variation of the surface reactivity under the influence of TSSs, which are consistent with Chen et al.'s theoretical predictions. In addition, our theoretical calculation indicates that the TSSs from $Bi_2Te_3$ can tunnel through a thin Pd layer and significantly enhance the surface reactivity of Pd.[8] It has also been well accepted that the highly desirable functionality in TI-based HER heterostructure is attributed to the TSSs as revealed by first-principles DFT calculations.[9,10] To the best of our knowledge, the experimental research on the use of a metal/TI or metal/TCI heterostructure as an electrocatalyst for HER with a desirable catalytic performance that is superior or comparable to that of a commercial Pt has not been realized prior to our work.

In this study, we report that a Pd/SnTe heterostructure could in fact offer such a desirable performance in HER. A Pd(20nm)/SnTe(70nm) heterostructure and a pure Pd(20nm) thin film were fabricated by molecular beam epitaxy (MBE) and electron beam evaporation. It was found that the Pd(20nm)/SnTe heterostructure shows much higher HER activity and turnover frequencies (TOFs) of per-Pd-site than those of the Pd(20nm) thin film. More importantly, the HER activity of the heterostructure is even superior to the Pt foil, which is attributed to our finding that electrons can be transferred from both the Pd surface and the adsorbed H atoms to the TSSs held by the SnTe (001) underlayer as revealed by our theoretical studies. As the exploration of the fascinating effects of the TSSs of topological materials is one of the major focuses in condensed matter physics research today, and the research in electrocatalytic water splitting is a highly important field in green energy generation, the findings of our study offer a new direction to optimize the performance of electrocatalysts based on topological quantum materials.

# Results

**Structural characterization and electrocatalytic performance of SnTe (001) thin films with various thicknesses**

As demonstrated in our previous report[6], the TSSs of SnTe can play an important positive role in the electrocatalytic activity, and the SnTe (001) sample exhibits a better electrocatalytic performance attributed to the larger electrochemical surface area than SnTe (111), thus we choose a SnTe (001) thin film as the TCI layer to provide the TSSs in the heterostructure. The cross-sectional high resolution transmission electron microscope (HRTEM) images of the SnTe (001) samples with various thicknesses are shown in Figure S1. The thicknesses of these SnTe thin films are determined to be 20 nm, 49 nm, 70 nm, 94 nm and 114 nm. It should be noted that one should only take these values as the approximate nominal thicknesses since these samples have domain structures at their surfaces, which will be addressed later. It is worthwhile to mention that as shown in Figure S1a and b, the 20 nm SnTe (001) thin film does not cover the surface of the GaAs substrate completely. The HRTEM images also show that all the SnTe thin films have a thin oxidized SnTe layer on top. Atomic force microscopy (AFM) images of the surfaces and the profile analyses of the five SnTe samples with nominal thicknesses of 20 nm, 49 nm, 70 nm, 94 nm and 114 nm are shown in Figure S2, and the details of the surface morphology and profile analyses are presented at the remark of Figure S2. As a summary, the growth of the SnTe (001) thin film on a GaAs substrate follows the island growth mode, and the 70 nm SnTe (001) sample exhibits the deepest depth of the holes and/or ditches and the largest surface area among the five SnTe (001) samples.

The results of the electrochemical catalytic HER behaviors of the five SnTe samples with various nominal thicknesses are presented in Figure S3, S4 and Table 1. The 70 nm SnTe (001) sample exhibits the lowest overpotentials ($\eta$) (198 mV) at a cathodic current density ($j$) of 10 mA cm$^{-2}$, the lowest Tafel slope of 49 mV/dec, the highest exchange current density ($j_0$) of 1.065 µA cm$^{-2}$ and the lowest charge transfer resistance ($R_{ct}$) of 19.84 Ω among the five SnTe samples. To obtain the estimated electrochemical surface areas (ECSAs) of these samples, the cyclic voltammetry (CV) measurements taken in the potential where no Faradic processes are observed were conducted. The capacitive current densities are plotted as a function of the scan rate as shown in Figure S4f. The resulting double-layer capacitances ($C_{dl}$) from the slopes of the lines shown in Figure S4f can be converted into the ECSA values. A detailed description of this method and our calculations for ECSAs of the five samples are provided in the remark of Figure S4. As can be seen, the 70 nm SnTe (001) thin film exhibits the highest estimated ECSA value of $32 \text{cm}^2_{ECSA}$ which is consistent with the AFM image analysis shown in Figure S2 regarding the deepest depth of the holes and ditches of the surface valley structure of this sample. We believe the highly efficient performance of the 70 nm SnTe (001) thin film is likely contributed by the fact that it has the largest active area among the five samples (as shown in Figure S2), which leads to a more efficient interaction between the catalytic electrode surface and the electrolyte.

Table 1 Comparison of the HER performances of the five SnTe (001) samples with nominal thicknesses of 20 nm, 49 nm, 70 nm, 94 nm, and 114 nm.

| Samples | Overpotential (mV vs. RHE) | Tafel slope (mv/dec) | Exchange current density $(j_0)$ ($\mu A\ cm^{-2}$) | $\text{Log}\ j_0$ ($A\ cm^{-2}$) | $R_{ct}$ (ohm) |
|---|---|---|---|---|---|
| SnTe 20 nm | 273 mV | 56 | 0.534 | -6.273 | 30.51 |
| SnTe 49 nm | 240 mV | 53 | 0.608 | -6.216 | 23.91 |
| SnTe 70 nm | 198 mV | 49 | 1.065 | -5.973 | 19.84 |
| SnTe 94 nm | 268 mV | 54 | 0.559 | -6.253 | 24.6 |
| SnTe 114 nm | 299 mV | 59 | 0.480 | -6.319 | 34.04 |

For the Pd/SnTe heterostructure, a seemingly straightforward approach to fabricate a Pd/SnTe heterostructure is to deposit a Pd layer directly onto a SnTe (001) thin film inside the MBE system. Such a sample was indeed fabricated and named Pd/SnTe-1. However, this approach was found to be less than optimal for the reasons to be addressed below. The high-resolution X-ray diffraction (HRXRD) profile of the Pd/SnTe-1 sample is shown in Figure S5. One can see that the two characteristic diffraction peaks locating at 28.24° and 58.395° match with the standard 2θ values of SnTe (002) and (004). The two characteristic diffraction peaks locating at 29.775° and 61.725° match with the standard 2θ values of $Pd_{71}Sn_7Te_{22}$ (202) and (404) (powder diffraction files #: 24-0821), indicated the *d*-spacing (*n*=1) is 2.9982 Å. The cross-sectional HRTEM images of the Pd/SnTe-1 sample are shown in Figure S6a-d. The energy dispersive spectroscopy (EDS) profiles of three different regions of the top layer (shown in Figure S6a) are displayed in Figure S7. One can see that the Pd, Sn and Te signals are all observed in the three EDS profiles, which indicates the top layer is a PdSnTe compound. Since the EDS technique is not a quantitative tool for measuring atomic concentrations, the exact ratios of the three elements in this compound cannot be concluded. Thus, we define this layer as $Pd_{1-x-y}Sn_xTe_y$. In addition, as shown in Figure S6b-d, the lattice spacing values along the *c*-axis for the three regions are measured to be 2.97 Å, 2.96 Å and 2.98 Å, which agree well with the *d*-spacing of 2.9982 Å calculated from the HRXRD profile. These findings demonstrate that *in situ* deposition of Pd directly on a SnTe layer will generate $Pd_{1-x-y}Sn_xTe_y$. The linear sweep voltammogram (LSV) of the Pd/SnTe-1 sample, the 70 nm SnTe (001) sample, and a Pt foil are shown in Figure S8. Although the $\eta$ at $j$ of 10 mA cm$^{-2}$ of the Pd/SnTe-1 sample is lower than that of the SnTe(70 nm), it is still higher than that of a Pt foil.

As reported by our previous work[6], SnTe (001) with partially oxidized surfaces can exhibit high activity, attributed to the enhanced charge transfer between H atoms and TSSs. Thus, for the Pd/SnTe heterostructures, a 70 nm SnTe layer was grown as the underlayer followed by the electron beam (E-beam) deposition of a 20 nm Pd overlayer *ex situ*, and an oxidized SnTe layer can be introduced between SnTe and Pd overlayer which could avoid the chemical reaction between Pd and SnTe. A pure Pd(20nm) thin film was also fabricated using the E-beam evaporation onto a GaAs substrate, which acts as a reference for comparison.

**Structural characterization of the Pd/SnTe heterostructure.**

Figure 1a and b show time-of-flight secondary ion mass spectrometry (ToF-SIMS) depth profiles of a typical Pd/SnTe heterostructure sample with positive and negative ions modes, respectively. The interfaces between the different layers are identified by analyzing the intensities change of characteristic ions (PdCs$^+$ and Pd$^-$ for Pd overlayer; CsSnO$^+$, O$^-$, SnO$_2^-$ and TeO$^-$ for the oxidized SnTe layer; CsSn$^+$, CsTe$^+$, SnTe$^-$, Te$^-$, and Sn$^-$ for SnTe thin film layer; CsGa$^+$, AsCs$^+$ and As$^-$ for the GaAs substrate) as shown by the vertical dash lines. The characteristic peaks of O$^-$, SnO$_2^-$, TeO$^-$ and CsSnO$^+$ demonstrate the existence of the thin SnTe oxidation layer between the SnTe thin film and the Pd overlayer. Besides, Pd$_{1-x-y}$Sn$_x$Te$_y$ compound ions including PdSn, PdTe, PdTe$_2$, *etc* were selected to check and no such compound ions are detected, which implies that Pd is unlikely to have a chemical reaction with SnTe and generate any Pd$_{1-x-y}$Sn$_x$Te$_y$ compound. Figure 1c displays the high resolution HRXRD 2θ-ω broad scan profile of the Pd/SnTe heterostructure sample. Two characteristic diffraction peaks locating at 28.215° and 58.34° match with the standard 2θ values of SnTe (002) and (004), respectively. Except for the diffraction peaks of SnTe (00*l*) and GaAs(00*l*), no other peaks of unexpected phases were observed. To obtain more crystal structural information of the surface of the Pd overlayer rather than the SnTe layer or the substrate, grazing incidence X-ray diffraction (GIXRD) was performed and the resulting 2θ scan profile is shown in Figure 1d. As can be seen, all the diffraction peaks can match well with the standard 2θ values of Pd, providing evidence that the Pd overlayer exists with the polycrystalline structure on the surface of the heterostructure.

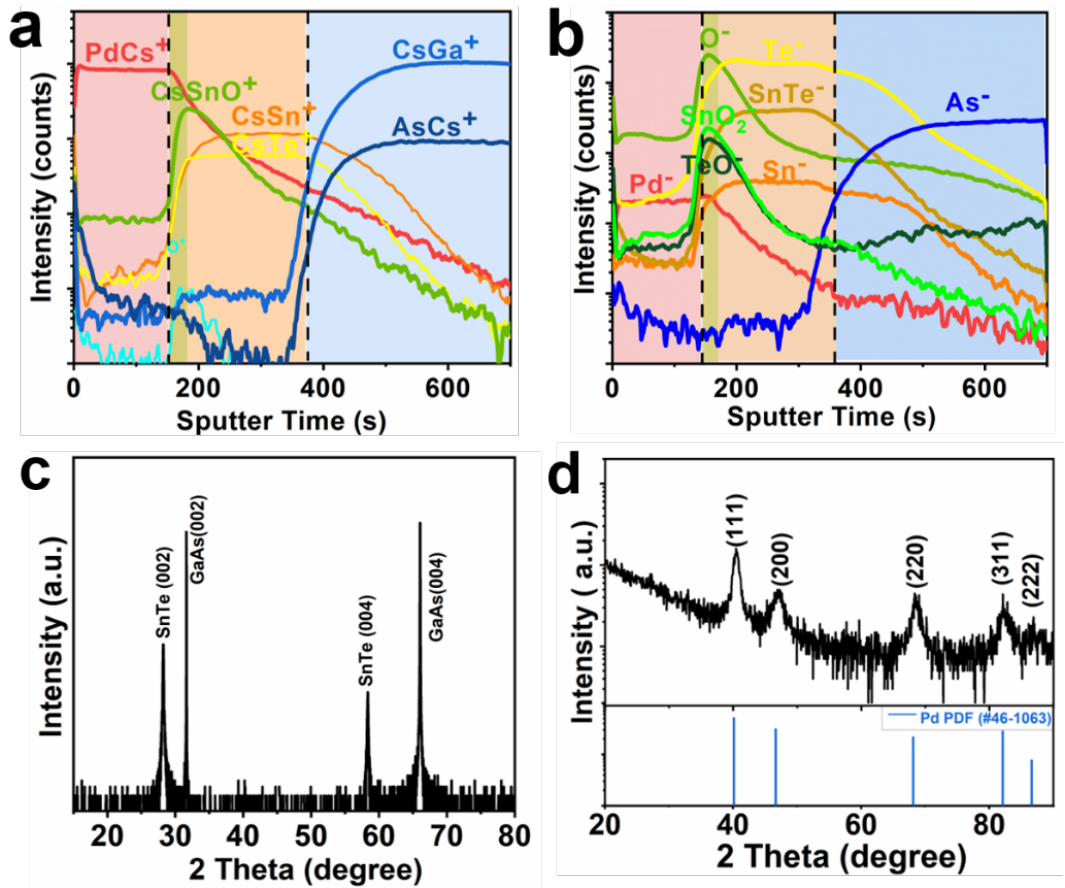

Figure 1. ToF-SIMS depth profiles for (a) positive and (b) negative ions measurements of a typical Pd/SnTe heterostructure sample. (c) High-resolution X-ray diffraction (HRXRD) 2θ-ω broad scan profile of the Pd/SnTe heterostructure sample. (d) Grazing incidence X-ray diffraction (GIXRD) profile of the Pd/SnTe heterostructure sample.

Figure 2a shows the cross-sectional high resolution HRTEM image of the Pd/SnTe heterostructure sample. It can be seen that the sample consists of a 1-2 nm SnTe oxidation layer which is consistent with the results of ToF-SIMS, and the thickness of Pd overlayer is about 20 nm. The corresponding fast Fourier transform (FFT) pattern of the Pd overlayer is shown in the inset of Figure 2a. Several diffraction points are indexed as Pd (111) and (220) with different orientations representing the polycrystalline structure of the Pd overlayer, which is consistent with the results of the GIXRD scan profile shown in Figure 1d. A cross-sectional schematic diagram of this heterostructure sample is displayed in Figure 2b. AFM images with an increasing magnification of the Pd(20nm)/SnTe heterostructure and the pure Pd(20nm) thin film are shown in Figure 2c, d, respectively. The surface morphology of the Pd(20nm)/SnTe heterostructure generally follows the valley structure of the SnTe (001) thin film as shown in Figure S2k, which provides larger active area than that of the surface of the pure Pd(20nm) sample.

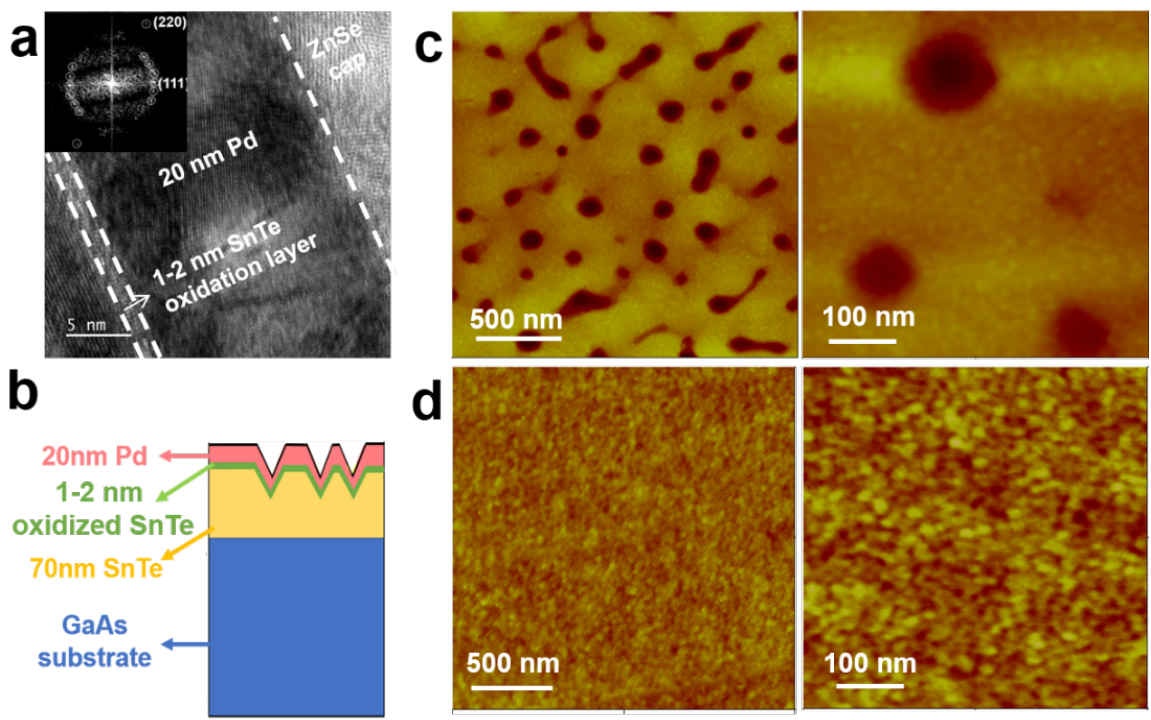

Figure 2. (a) Cross-sectional HRTEM image of the Pd(20nm)/SnTe heterostructure. (b) A cross-sectional schematic diagram of this heterostructure. AFM images with an increasing magnification of (c) the Pd(20nm)/SnTe heterostructure and (d) the pure Pd(20nm) sample.

**Electrocatalytic performance of the Pd/SnTe heterostructure.**

The HER electrocatalytic behaviors of the Pd(20nm)/SnTe heterostructure and the pure Pd(20nm) sample are shown in Figure 3. All the corresponding data are listed in Table 2. Figure 3a displays the LSV polarization curves of the Pd(20nm)/SnTe heterostructure, the Pd(20nm) sample, the 70 nm SnTe (001) thin film which is used in the heterostructure, and a commercial Pt foil. The $\eta$ of the Pd(20nm)/SnTe heterostructure at a cathodic $j$ of 10 mA cm$^{-2}$ is 86 mV, surpassing that of the Pd(20nm) sample (140 mV) and the commercial Pt foil (94 mV). The corresponding Tafel plots of the Pd(20nm)/SnTe heterostructure, the pure Pd(20nm) sample and the commercial Pt are displayed in Figure 3b and the linear portions of the plots are fitted to the Tafel equation ($\eta = b\log j + a$, where $j$ is the current density and $b$ is the Tafel slope). As shown in Figure 3b, the Tafel slope of the Pt foil as a reference is 29.83 mV/dec which is consistent with the expected value of the known HER mechanism of Pt. The fact that the Pd(20nm)/SnTe heterostructure has a lower Tafel slope of 29.14 mV than that of the pure Pd(20nm) sample (31.45 mV/dec) and the Pt foil (29.83 mV/dec), revealing a better HER kinetics of the Pd(20nm)/SnTe heterostructure. The $j_0$ that reflects the inherent HER kinetics was determined by the extrapolated x-intercepts ($\eta$=0) of the fitted lines. The $j_0$ of the Pd(20nm)/SnTe heterostructure (129.1 μA cm$^{-2}$) is higher than that of Pt foil (116.4 μA cm$^{-2}$) and over twice of that of the pure Pd(20nm) sample (57.28 μA cm$^{-2}$), which is consistent with the observation that its $\eta$ and Tafel slope are lower than those of the pure Pd(20nm) sample and Pt foil. The corresponding log $j_0$(A/cm$^2$) values are also given in Table 2. Figure 3c shows the results of the electrochemical impedance spectroscopy (EIS) measurements performed on the Pd(20nm)/SnTe heterostructure and the pure Pd(20nm) sample. By applying a simplified Randles circuit model as shown in the inset of Figure 3c, we can derive the $R_{ct}$ of them. The results are shown in Table 2 and it can be seen that the Pd(20nm)/SnTe heterostructure exhibits a lower $R_{ct}$ of 3.517 Ω than that of the pure Pd(20nm) sample (10.01 Ω), suggesting a more efficient charge transfer between the catalysts surface and electrolyte in the former, which is also consistent with the observation that its $\eta$ and Tafel slope are lower and $j_0$ is higher than those of the pure Pd(20nm) sample. Figure 3d shows the polarization curves recorded before and after 1000 cycles of CV treatments to evaluate the long-term stability. There is an observed potential increase of only 11.14 mV at a cathodic current density $j$ = 10 mA cm$^{-2}$ after the CV treatments, representing that the electrochemical HER process does not cause significant loss in the electrocatalytic performance of the Pd(20nm)/SnTe heterostructure. This slight degradation may arise from that some active components may be exfoliated during the CV treatments due to the disturbance from the evolution of H$_2$ bubbles generated on the Pd(20nm)/SnTe cathode.[11] We partially attribute the excellent performance in HER of the Pd(20nm)/SnTe heterostructure which is superior to Pd(20nm) to the rougher valley structure of the surface morphology of the Pd(20nm)/SnTe heterostructure that providing larger electrochemical surface area as revealed by AFM images.

To identify the inherent role of the SnTe layer besides offering a larger surface area for the highly efficient electrocatalytic performance of the Pd(20nm)/SnTe heterostructure, the intrinsic activities of the heterostructure and the pure Pd(20nm) sample are evaluated by extracting the turnover frequency (TOF), which requires one to determine the ratio between the number of totally hydrogen turnover per cm$^2$ and the number of active sites per cm$^2$. Figure 3e, f show the results of CV measurements with different scan rates taken in the potential where no Faradic processes are observed, with the aim to obtain the estimated ECSAs. The capacitive current densities (($j_{anodic}$-$j_{cathodic}$)/2 taken at the potential values of the dash lines of Figure 3e and f) are plotted as a function of the scan rates as shown in Figure 3g. The slopes representing the double-layer capacitance ($C_{dl}$) were extracted to be 0.505 mF cm$^{-2}$ and 0.320 mF cm$^{-2}$ for the Pd(20nm)/SnTe heterostructure and the pure Pd(20nm) sample, respectively. The ECSA were estimated from the $C_{dl}$ values as shown in Supplementary note 1. The estimated ECSA values are consistent with the AFM image analysis shown in Figure 2c, d regarding the degree of surface roughness that the Pd(20nm)/SnTe heterostructure can exhibit larger electrochemical surface area than the pure Pd(20nm) sample. It is well known that the intrinsic electrocatalytic activity of a catalytic surface can be evaluated by extracting the TOF, which requires one to determine the ratio between the number of total hydrogen turnover per cm$^2$ and the number of active sites per cm$^2$.[12] A detailed description of this method and our calculations are provided in Supplementary note 1. Figure 3h shows the plots of TOF values versus the potentials for the Pd(20nm)/SnTe heterostructure and the pure Pd(20nm) sample. Apparently, the TOF values of the Pd(20nm)/SnTe heterostructure are much larger than that of the Pd(20nm) sample, indicating that the per-Pd-site of the Pd(20nm)/SnTe heterostructures likely enjoys a higher intrinsic activity than that of the pure Pd(20nm) sample. The TOF value of the heterostructures is more than twice higher than that of the Pd(20nm) sample at a potential of 0.2 V.

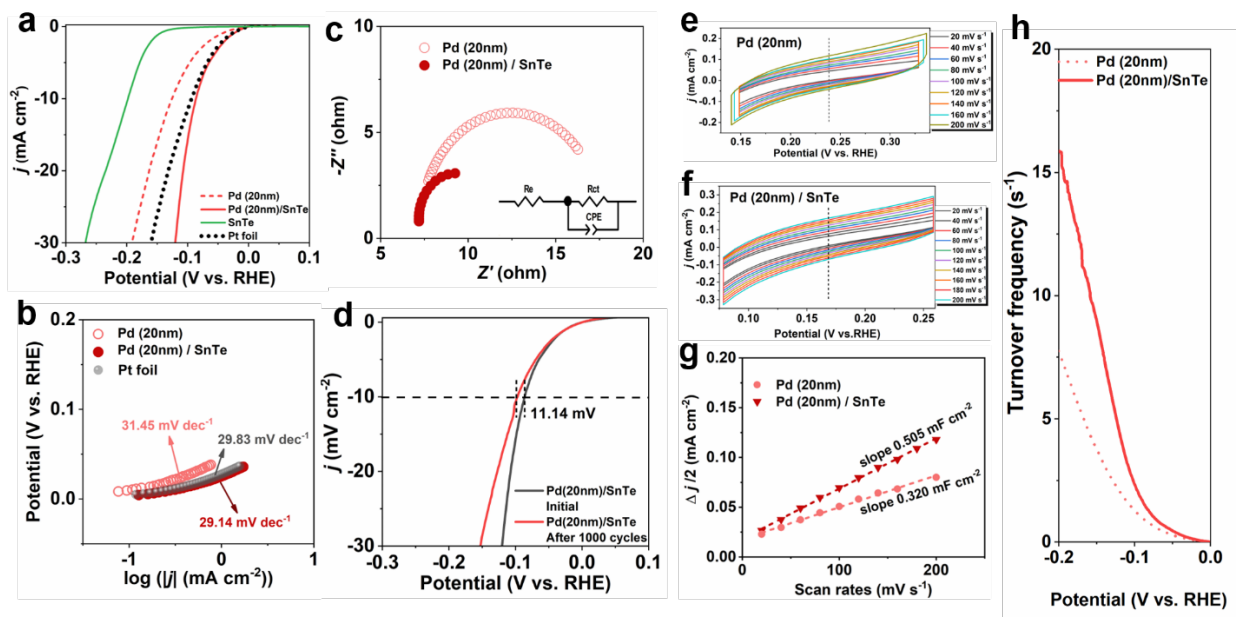

Figure 3. (a) Polarization curves (*iR*-corrected) of the pure Pd(20nm) sample, the Pd(20nm)/SnTe heterostructure, the 70 nm SnTe (001) thin film sample, the Pd(20nm)/SnTe heterostructure and a commercial Pt foil. (b) Corresponding Tafel plots of the pure Pd(20nm) sample, the Pd(20nm)/SnTe heterostructure, and the commercial Pt. (c) Nyquist plots of the pure Pd(20nm) sample and the Pd(20nm)/SnTe heterostructure. Inset: schematic drawing of the simplified Randles circuit, the equivalent circuit model used for fitting the EIS response of HER, where $R_s$ is the electrolyte resistance and $R_{ct}$ denotes the charge-transfer resistance. (d) Polarization curves of the Pd(20nm)/SnTe sample recorded before and after 1000 cycles of cyclic voltammetry (CV) using accelerated degradation tests (scan rate = 100 mV s$^{-1}$). CV curves with different scan rates in the potential ranges where no faradic processes are observed for the (e) pure Pd(20nm) sample and (f) the Pd(20nm)/70SnTe heterostructure. (g) Linear fits of the half capacitive currents as a function of scan rates for the extraction of $C_{dl}$ values of the pure Pd(20nm) sample and the Pd(20nm)/SnTe heterostructure. (h) Turnover frequency (TOF) plots against the potentials of the pure Pd(20nm) sample and the Pd(20nm)/SnTe heterostructure.

Table 2. Comparison of the HER performances among the Pd(20nm)/SnTe heterostructure, the pure Pd(20nm) sample and the Pt foil.

| Samples | Overpotential (V vs. RHE) | Tafel slope (mv/dec) | $j_0$ ($\mu A\ cm^{-2}$) | Log $j_0$ ($A\ cm^{-2}$) | $R_{ct}$ ($\Omega$) |
|---|---|---|---|---|---|
| Pd(20nm)/SnTe | 0.086 | 29.14 | 129.1 | -3.889 | 3.517 |
| Pd(20nm) | 0.140 | 31.45 | 57.28 | -4.242 | 10.01 |
| Pt foil | 0.094 | 29.83 | 116.4 | -3.934 | N.A. |

**Theoretical studies**

We conducted DFT calculations to study the mechanism behind the high HER performance of Pd-covered SnTe thin films (Pd/SnTe heterostructures) found in experiments. As shown in previous studies[4], the free energy of hydrogen adsorption $\Delta G_H$ is a good indicator for the HER activity of various electrocatalysts. The optimal value of $\Delta G_H$ is ~0 eV, as the H atom is neither too attractive nor too repulsive to the catalyst surface.

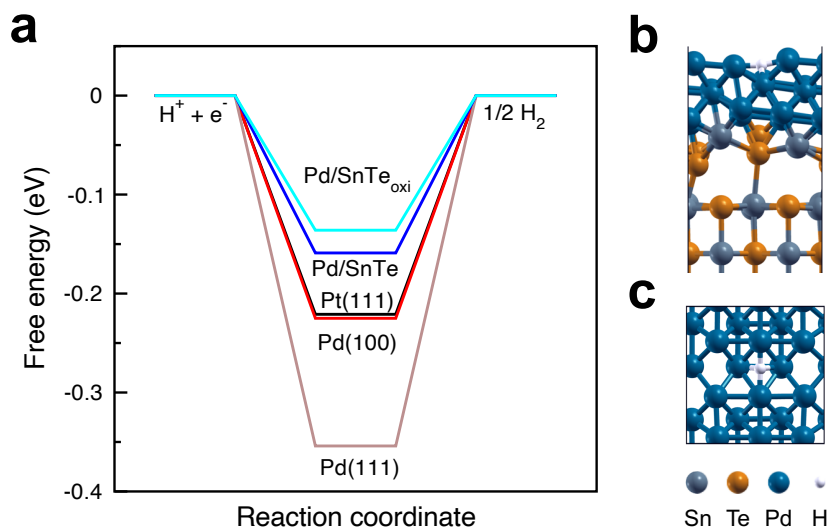

Figure 4. (a) Free energies of hydrogen adsorption ($\Delta G_H$) on the Pd (111) surface, the Pd (100) surface, the Pt (111) surface, the Pd/SnTe (001) heterostructure, and the partially oxidized heterostructure, Pd/SnTe$_{oxi}$. The hydrogen coverage is 1/9 ML. (b) Side view and (c) top view of the Pd/SnTe heterostructure.

First, we tested a variety of hydrogen adsorption sites on the surface of the Pd/SnTe heterostructure. We found that $\Delta G_H$ of the Pd/SnTe heterostructure is generally closer to 0 eV than that of the pristine Pd surface on the most active adsorption sites. Our experimental measurements suggested that the SnTe (001) layers covered by the Pd slabs are not clean, but partially oxidized, so we also considered the partial oxidation at the interface between Pd and SnTe (Pd/SnTe$_{oxi}$). We calculated the formation energies ($E_f$) of the Pd overlayers grown on the pure and oxidized SnTe slabs (see Supplementary note 4). The Pd/SnTe$_{oxi}$ heterostructure has lower $E_f$ than the Pd/SnTe heterostructure by 0.43 eV, indicating that the Pd/SnTe$_{oxi}$ heterostructure is more energetically favored. Figure 4a shows that $\Delta G_H$ of the Pd/SnTe$_{oxi}$ slab with the partially oxidized interface (-0.14 eV) is slightly closer to 0 eV than that of the Pd/SnTe slab without oxidation (-0.16 eV), suggesting that the partial oxidation in Pd/SnTe$_{oxi}$ could provide slightly better HER performance. Overall, our theoretical calculations are consistent with experimental results, indicating that both Pd/SnTe and Pd/SnTe$_{oxi}$ heterostructures have higher HER performance than pure Pd or Pt (see Figure 4a).

In the following, we study the role of the TSSs of SnTe (001) without oxidation in the heterostructures on the electron transfer processes, since even though the partial oxidation may affect the structural stability, however, the roles of TSSs in both heterostructures (with or without partial oxidation) should be similar. Figure S10 displays the band structure of Pd/SnTe heterostructure, which shows that the

TSSs can exist on the heterostructure. We applied the Löwdin population analysis[13] to calculate the change of atomic charges after removing the SnTe substrate, and found that 0.0108 electrons and 0.0975 electrons transfer from the adsorbed H atom and the H-hosting Pd atoms, respectively, to the SnTe slab. Our DFT calculations suggest that the TSSs of the SnTe upper surface in the Pd/SnTe heterostructure work as an electron bath to accept electrons transferred from the adsorbed H atoms. Similar electron transfer processes are also found in the partially oxidized and Sn-vacancy-containing SnTe (001) surface[6] and other topological materials[5,7,9]. The interaction between H atom and a pure Pd surface is too attractive, resulting in a large negative $\Delta G_H$ value as shown in Figure 4, which is also supported by previous volcano plot calculations.[4,14,15] Our study shows that both the adsorbed H atoms and the Pd surface at the Pd/SnTe heterostructure are electron-deficient, which work together to weaken the H-Pd bond, thus enhancing the HER performance.[16,17]

Many previous studies applied the d-band theory to explain the electrocatalytic activity of transition metal heterostructures, which claims that the downshift of the d-band center with respect to the fermi level leads to weaker hydrogen adsorption.[15,18–21] In our case, however, enhancement of electrocatalytic performance cannot be explained by the d-band theory, highlighting the unconventional role of the SnTe substrate. As shown in Figure 5, the d-band center of Pd/SnTe heterostructure surface moves up to the Fermi level by 0.08 eV as compared to that of pure Pd (100) surface. According to the linear relation between the hydrogen adsorption energy and the d-band center obtained by Kitchin et al[22], if the d-band center had a dominating effect on the hydrogen adsorption process, $\Delta G_H$ should decrease by 0.02 eV after adding the SnTe substrate. However, our DFT calculations show that $\Delta G_H$ increases by 0.07 eV (see Figure 4), which suggests that the conventional d-band theory does not work here. Our theoretical results suggest that the high HER performance of Pd/SnTe heterostructures is not due to the d-band shift, but the weakened H-Pd binding caused by the electron transfer from the adsorbed H atom and the Pd surface to the TSSs of the SnTe (001) underlayer.

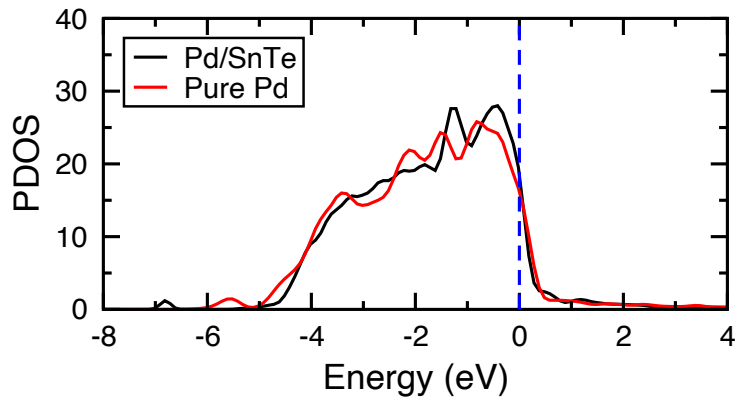

Figure 5. Projected density of states (PDOS) of d-band states in the surfaces of Pd/SnTe heterostructure (black line) and pure Pd (100) slabs (red line). After adding the SnTe substrate, the d-band center moves closer to the Fermi level by 0.08 eV. The Fermi level is set to zero.

In our experiments, we tested Pd/SnTe heterostructures with 5, 8, 14, 20, 30, 50 nm thick Pd overlayers in the Pd/SnTe heterostructures and found that the 20 nm Pd/SnTe heterostructure has the best performance in HER. This experimental observation may come from two causes: 1) based on our AFM and TEM studies shown in Figure S11 and S12, Pd layers ≤ 8 nm are discontinuous; 2) the experimental Pd overlayers are polycrystalline, while in our calculations, we used crystalline Pd (100) slab. It is worthwhile to mention that the mechanism revealed by our theoretical study should be valid for both crystalline and polycrystalline Pd. If a thinner and continuous crystalline Pd (100) overlayer can be deposited on SnTe (001), we expect even higher HER performance and the catalyst can be more cost effective because less Pd is required, which is a promising direction to explore in the future.

## Conclusion

In this work, SnTe (001) thin films with various thicknesses were fabricated by the MBE technique. It was found that the 70 nm SnTe (001) thin film exhibits the best HER performances among them, which is attributed to its largest active areas as revealed by AFM imaging and the estimated ECSA from the $C_{dl}$ measurements using CV. A Pd(20nm)/SnTe(70nm) heterostructure was grown with the Pd overlayer deposited *ex situ* using electron-beam evaporation onto the MBE-grown SnTe layer, which was proved to be able to avoid $Pd_{1-x-y}Sn_xTe_y$ compound formation that occurs if the heterostructure is grown *in situ* using the MBE technique alone. ToF-SIMS and HRTEM demonstrate that the heterostructure consists of a 1-2 nm oxidized SnTe layer at the interface between the Pd and SnTe layers. GIXRD and cross-sectional HRTEM provide the evidence that the Pd overlayer exists with the polycrystalline structure. The HER characterizations reveal that the $\eta$ of the heterostructure at $j$ of 10 mA cm$^{-2}$ and its Tafel slope are lower than those of the pure Pd(20nm) sample and a commercial Pt foil. The $j_0$ of the heterostructure is higher than that of the Pt foil and over twice of that of the pure Pd(20nm) sample and the heterostructure also exhibits a lower $R_{ct}$ than that of the Pd(20nm) sample. By conducting LSV and CV measurements to reveal their intrinsic activities, it was found that the heterostructure enjoys a much higher intrinsic activity per Pd site than that of the Pd(20nm) sample based on the extracted TOF values. The TOF value of the heterostructure is more than twice higher than that of the Pd(20nm) sample at a potential of 0.2 V. Our theoretical studies reveal that electrons transfer from both the Pd surface and the adsorbed H atoms to the TSSs in the SnTe (001) underlayer, resulting in weaker Pd-H binding and more favorable $\Delta G_H$ values for HER. This work demonstrates that the TSSs of the SnTe (001) surfaces play an important role in enhancing the HER activity of Pd and even exceeding that of a Pt foil, which suggests a new approach to optimize the HER performance of electrocatalysts using the interesting properties of the TSSs in topological quantum materials.

# Methods

**Sample preparation**

*SnTe (001) thin films with various thicknesses*

All the SnTe (001) thin film samples used in this work were fabricated on n+ GaAs (001) substrates at a substrate temperature of 222 °C by a VG-V80H MBE system equipped with *in situ* RHEED. The growth processes were performed in an UHV chamber with a basic pressure better than $1.0 \times 10^{-9}$ Torr. SnTe (001) thin films with thicknesses of 20 nm, 49 nm, 70 nm, 94 nm and 114 nm were grown at the same growth rate and the thickness of SnTe was controlled by the growth time.

*Pd/SnTe-1 sample*

The growth of Pd in Pd/SnTe-1 was performed at a substrate temperature of 30 °C after the growth of a 70 nm SnTe (001) thin film in the MBE system.

*Pd thin film and Pd/SnTe heterostructure*

A pure Pd thin film of 20 nm was deposited on an epi-ready n+ GaAs (001) substrate using an IVS EB-500M E-beam evaporator. After exposed in air for a short time, a piece of an MBE-grown 70 nm SnTe thin film sample was transferred to the same evaporator for a further deposition of Pd with a thickness of 20 nm on top. All the Pd depositions were handled with the substrate temperature at room temperature and with the same deposition rate of 1 Å/s. The base vacuum of the evaporator before the deposition work is better than $3 \times 10^{-7}$ Torr and the deposition processes were performed with a base pressure better than $1.6 \times 10^{-6}$ Torr.

## Material structural characterization

The samples were characterized by HRXRD using a PANalytical multipurpose diffractometer with Cu K$\alpha_1$ X-rays at a wavelength of 1.54056 Å. For GIXRD, a PANalytical multipurpose diffractometer was used with the incident beam kept at an angle of 0.5° in order to reduce both the penetration depth of incoming X-rays and the bulk diffuse scatting. HRTEM imaging was handled using a JEOL JEM-2010F with an acceleration voltage of 200 kV, and AFM profiling was carried out using a Dimension 3100 with a NanoScope IIIa controller (Digital Instruments) via the tapping mode. ToF-SIMS depth profiling were carried out using a dual-beam ToF-SIMS V (ION-TOF GmbH) system with a 25 keV Bi$^{3+}$ primary ion beam for analysis steps and 3 keV Cs$^+$ ion beams for sputtering steps. The depth profiling is built by alternating sputtering and analysis steps on typically $500 \times 500$ μm$^2$ sputter areas and $100 \times 100$ μm$^2$ analysis areas, respectively. The resulting primary ion and sputtering ion doses density are less than $1.43 \times 10^{13}$ ions/cm$^2$ and $5.73 \times 10^{16}$ ions/cm$^2$ respectively.

## Electrochemical measurements for hydrogen evolution reaction

All the electrochemical measurements were performed in a standard three-electrode electrolyzer connected to a CHI 660E electrochemical workstation (CH Instruments), using samples we studied in this work as the working electrode. A carbon rod was used as a counter electrode and a standard Ag/AgCl electrode (saturated KCl solution) served as the reference electrode. The electrolyte was a 0.5 M H$_2$SO$_4$ solution (degassed by N$_2$ with purity ~99.995%). In constructing the working electrode, a piece of the as-grown sample is connected to a poly-ether-ether-ketone (PEEK) electrode via a conductive glassy carbon clip, one side of the clip touches the back side of the GaAs substrate and the other side connects a small part of the sample surface to the PEEK electrode. As mentioned in our previous work[5], the GaAs substrate has negligible HER activity. LSV was performed using a scan rate of 5 mV s$^{-1}$. EIS measurements were carried out at open circuit potential over a frequency range of $10^6$ to 0.01 Hz with a perturbation voltage amplitude of 5 mV. The impedance data were fitted to a simplified Randles circuit to extract series and charge-transfer resistances. The electrochemical double layer capacitance of the electrode for estimation of electrochemical surface area was determined by cyclic voltammetry at different scan rates (from 20 to 200 mV s$^{-1}$) in the potential ranges where no faradic processes are observed. The TOF values of the samples were calculated according to the literature procedure[12]. All data presented were *iR* corrected. The potential values shown in later figures or tables were defined with respect to the reversible hydrogen electrode.

**DFT calculations**

We conducted the DFT calculations using plane-wave basis sets and the Perdew-Burke-Ernzerhof (PBE) exchange-correlation functional[24] implemented in the Quantum Espresso package (v.6.1).[25] The kinetic energy cutoff of plane waves is 50 Ry. We used the ONCV pseudopotentials for the H, Sn, Te and Pd atoms, and the ultrasoft pseudopotential for the O atom.[26–28] We applied the experimental lattice constants of 6.328 Å for the SnTe rock-salt structure[29–31] and 3.89 Å for the Pd face-centered cubic structure[32]. In order to reduce the lattice mismatch between SnTe (001) and Pd (100), we adopted a 2 × 2 supercell of SnTe (001) and a 3 × 3 supercell of Pd (100) so that the Pd (100) overlayer has a tensile strain of 8.4%. We tested the Pd (100) slab under the tensile strain from 0 to 8.4% and found that $\Delta G_H$ at the bridge adsorption site remains unchanged. In structural relaxations, the Pd/SnTe heterostructure consists of 3 ML Pd (100) and 6 ML SnTe (001) slabs, and the bottom 4 ML SnTe is fixed. In electronic structure calculations, we increased the thickness of the SnTe slab to 10 ML and included spin-orbit coupling to identify the gapless TSSs. We used 12 Å vacuum to avoid interactions between neighboring slabs with periodic boundary conditions. We used the 3 × 3 × 1 Monkhorst-Pack k-point mesh[33] in structural relaxations and 4 × 4 × 1 in electronic structure calculations.

The free energy of adsorption $\Delta G_H$ was calculated as:

$$\Delta G_H = \Delta E_H + \Delta E_{ZPE} - T\Delta S_H \quad (1).$$

Here $\Delta E_H$ represents the hydrogen adsorption energy calculated by:

$$\Delta E_H = E[slab + H^{ad}] - E[slab] - \frac{1}{2}E[H_2] \quad (2)$$

where $E[slab + H^{ad}]$ is the total energy of the Pd/SnTe or Pd/SnTe$_{oxi}$ heterostructure with one adsorbed H atom, $E[slab]$ is the energy of the Pd/SnTe or Pd/SnTe$_{oxi}$ heterostructure without the H adsorption, and

$E[H_2]$ is the energy of hydrogen molecule in the gas phase. $\Delta E_{ZPE}$ represents the change of the zero-point energy in the H adsorption, which was calculated by:

$$\Delta E_{ZPE} = E_{ZPE}[H^{ad}] - \frac{1}{2} E_{ZPE}[H_2] \quad (3)$$

where $E_{ZPE}[H^{ad}]$ is the zero-point energy of one H atom adsorbed on the Pd/SnTe or Pd/SnTe$_{oxi}$ heterostructure, and $E_{ZPE}[H_2]$ is the zero-point energy of hydrogen molecule in the gas phase. $\Delta S_H$ is calculated as $-\frac{1}{2} S^0_{H_2}$, where $S^0_{H_2}$ is the entropy of hydrogen molecule in the gas phase at the standard condition (130.68 J·$mol^{-1}$·$K^{-1}$ at $T = 298\ K$ and $p = 1\ bar$).[34] By combining the last two terms in equation (1), $\Delta G_H$ is calculated simply as $\Delta G_H = \Delta E_H + 0.25\ eV$.

## Author Contributions

Q.Q. and B.L. contributed equally to this work. Q.Q. and I.K.S. initiated this study and further designed the experiments; Q.Q. and W.S.L carried out the sample synthesis, conducted the structural characterizations and electrochemical measurements; B.L. and D.P. carried out the theoretical calculations; Q.Q., I.K.S., B.L., and D.P. wrote the manuscript. All authors performed the data analysis and discussions.

## Conflicts of interest

The authors declare that they have no known competing financial interests or personal relationships that could have appeared to influence the work reported in this paper.

## Acknowledgements

This research was funded by the Research Grants Council of the Hong Kong Special Administrative Region, China, under Grant Numbers 16304515, 16308020, C6025-19G, C6011-20G and William Mong Institute of Nano Science and Technology under Project Number WMINST19SC07. D.P. acknowledges support from the Croucher Foundation through the Croucher Innovation Award and the Hetao Shenzhen/Hong Kong Innovation and Technology Cooperation (HZQB-KCZYB-2020083).

Supplementary information

# Highly active hydrogen evolution facilitated by topological surface states on a Pd/SnTe metal/topological crystalline insulator heterostructure

*Qing Qu[1, 2†], Bin Liu[1†], Wing Sum Lau[1], Ding Pan[1,3,4*] and Iam Keong Sou[1,2*]*

[1]Department of Physics, The Hong Kong University of Science and Technology, Clear Water Bay, Hong Kong, China.

[2]William Mong Institute of Nano Science and Technology, The Hong Kong University of Science and Technology, Hong Kong, China.

[3]Department of Chemistry, The Hong Kong University of Science and Technology, Hong Kong, China.

[4] HKUST Shenzhen-Hong Kong Collaborative Innovation Research Institute, Shenzhen, China.

[†] These authors contributed equally to this work.

\* Corresponding authors

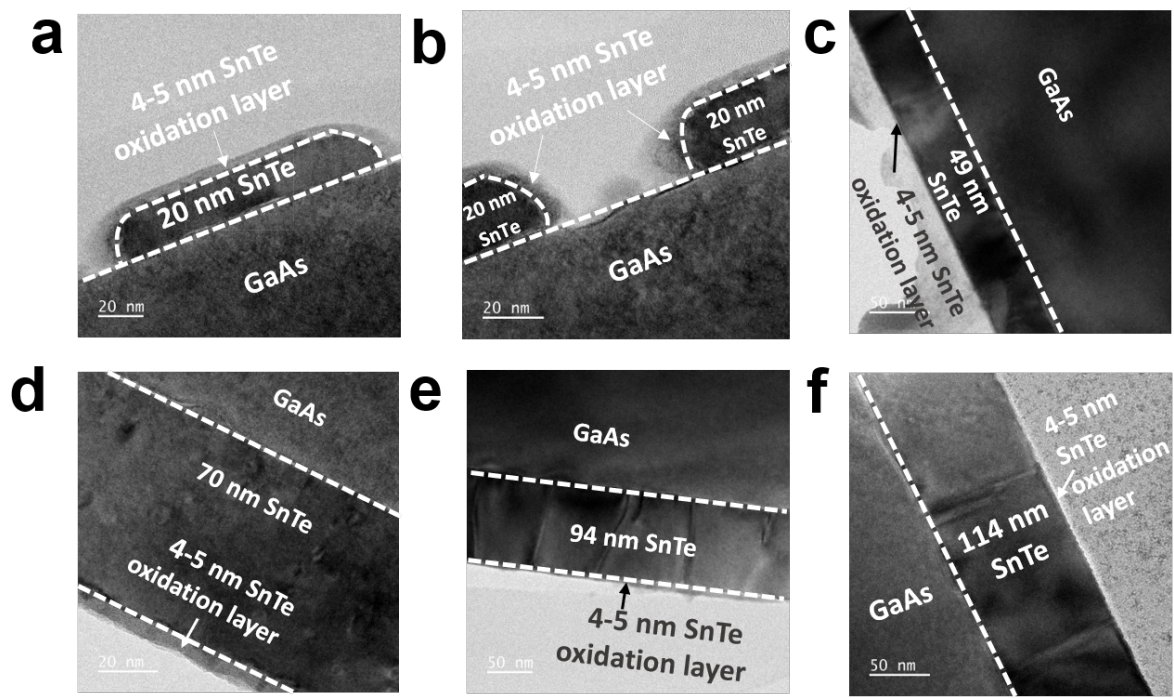

Figure S1 Cross-sectional transmission electron microscopy (TEM) images of the SnTe (001) samples with thicknesses of (a, b) 20 nm, (c) 49 nm, (d) 70 nm, (e) 94 nm and (f) 114 nm.

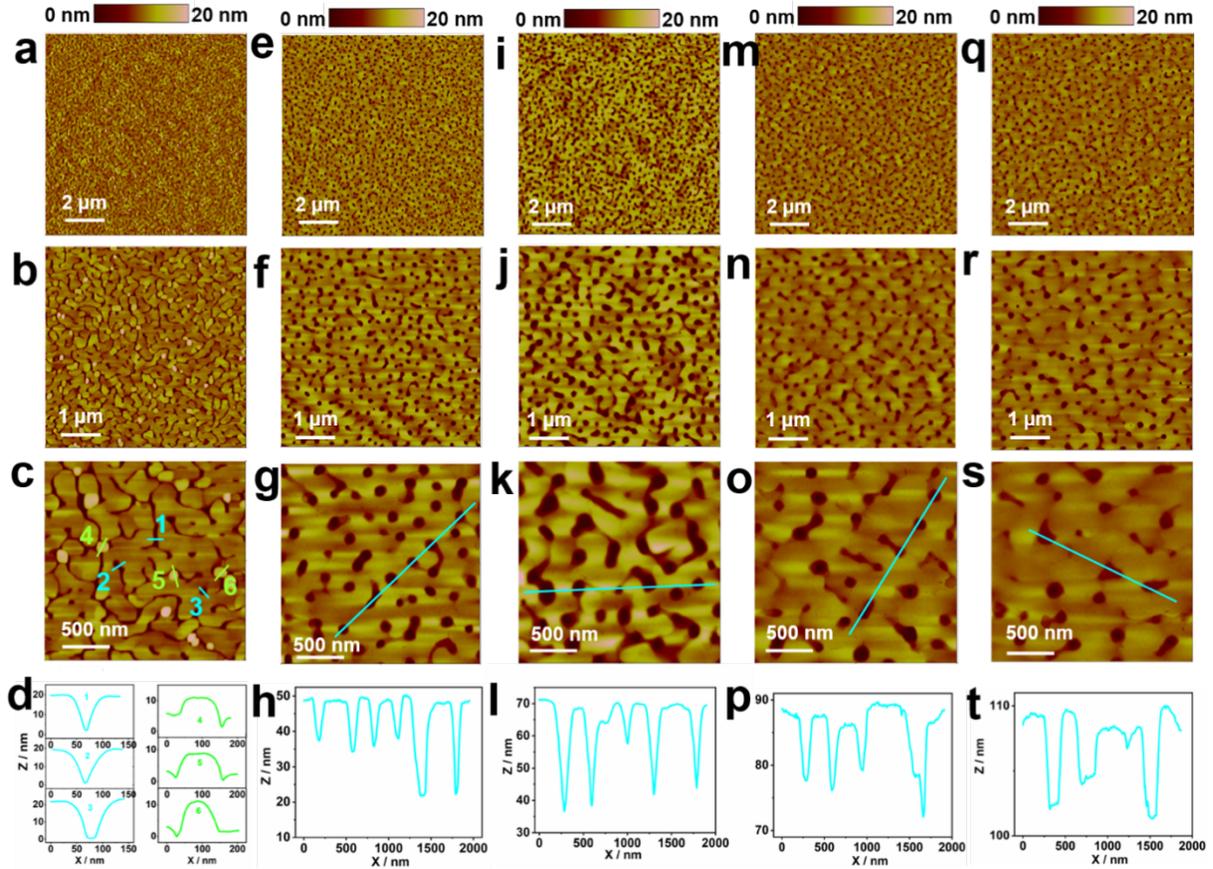

Figure S2 Atomic force microscopy (AFM) images of the five SnTe (001) samples with nominal thicknesses of (a-c) 20 nm, (e-g) 49 nm, (i-k) 70 nm, (m-o) 94 nm, and (q-s) 114 nm. The profile analyses of (c, g, k, o, s) along the green or blue lines are presented in (d, h, l, p, t) respectively.

Remark of Figure S2:

Figure S2a-c display the AFM images of the surface of the 20 nm SnTe sample with increasing magnification. As shown in Figure S2c, flat islands between narrow ditches and some perpendicular nanorods can be clearly observed at the surface of this sample. Such a surface morphology of island growth is similar to the observation by Ishikawa et al.[1] The profile analyses carried out for some typical ditches and nanorods are shown in Figure S2d. It was found that the depths of the ditches are around 20 nm which corresponds to the thickness of the islands observed in the TEM images shown in Figure S1a-b, thus the bottom parts of the ditches should be the exposed surface of the GaAs substrates. The heights of the nanorods are around 30 nm and the diameter of the nanorods are around 100 nm. When the nominal thickness of SnTe reaches 49 nm, as shown in Figure S2e-g, several preferred islands seemed to coalesce to form larger contiguous islands with holes and wider ditches. Figure S2h shows the profile analysis along

the blue line in Figure S2g and the depth of the holes and the ditches are ranged from 12 nm to 28 nm. Figure S2i-k are the AFM images of the nominal thickness of the 70 nm SnTe thin film and Figure S2l is the profile analysis carried out for some typical holes and ditches along the blue line in Figure S2k. It displays that the surface of 70 nm SnTe thin film exhibits a valley structure with holes and ditches with a depth of 30 nm of holes and ditches, which is deeper than that of the 49 nm SnTe (001) sample. When the thickness of SnTe reaches 94 nm, as shown in Figure S2m-o, several neighboring domains seem to be getting merged with each other forming larger islands with smaller holes and narrower ditches. Figure S2p shows the profile analysis along the blue line in Figure S2o. It can be seen that the depth of these holes and ditches is less than 20 nm, which seems to provide evidence that the further growth of SnTe (001) occurred at the places with holes and ditches between the islands that have been almost fully developed. Figure S2q-s show the AFM images of the surface of the 110 nm SnTe (001) sample in which the continuous domains are even larger with smaller holes and narrower ditches than those of the 94 nm SnTe (001) sample. Figure S2t shows the profile analysis along the blue lines in Figure S2s and the depth of the holes and ditches became shallower than 10 nm, revealing that the further growth of SnTe follows the growth mode of 94 nm as described above.

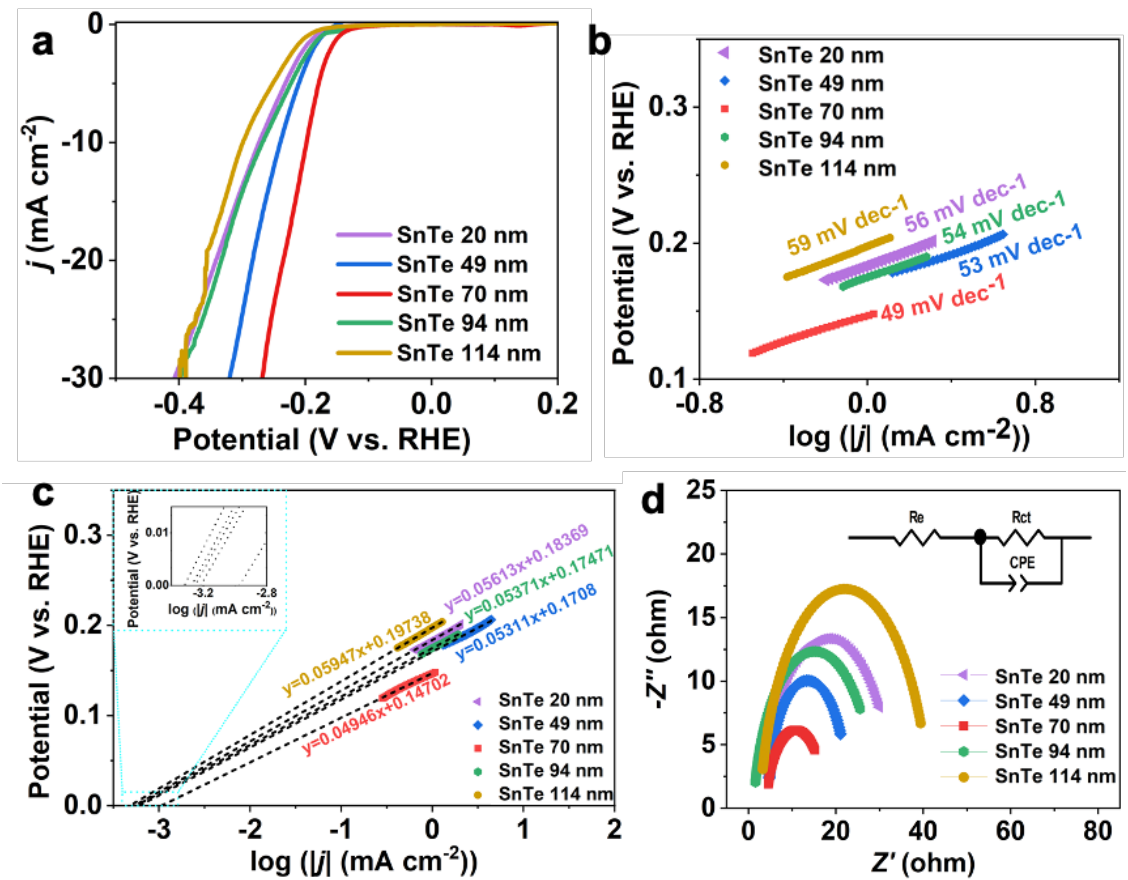

Figure S3 HER electrocatalytic performances of the five SnTe (001) samples with nominal thicknesses of 20 nm, 49 nm, 70 nm, 94 nm and 114 nm. (a) Polarization curves (*iR*-corrected) of the five SnTe samples. (b) Corresponding Tafel plots of the materials in (a). (c) Exchange current densities ($j_0$) for the materials in (a) can be derived from the x-intercepts of the fitted dash lines extrapolated from the Tafel plots shown in colorful scatters. (d) Nyquist plots of the materials in (a). The inset is the simplified Randles circuit model used for fitting the EIS response of HER.

Remark of Figure S3:

Figure S3a displays the linear sweep voltammogram (LSV) curves of the five SnTe samples. The overpotentials ($\eta$) at a cathodic current density ($j$) of 10 mA cm$^{-2}$ of the five SnTe samples ranged within 198-299 mV, and the 70 nm SnTe (001) sample exhibits the lowest $\eta$ (198 mV) among them. The kinetics of the activity of catalytic materials during the HER process can be revealed by the Tafel slope. The corresponding Tafel plots of the five SnTe (001) samples are displayed in Figure S3b. The linear portions of these plots are fitted to the Tafel equation ($\eta = b \log j + a$, where $j$ is the current density and $b$ is the Tafel slope; the fitting equations are displayed in Figure S3c), yielding a range from 49 to 59 mV/ dec for the five SnTe samples. For the 70 nm SnTe (001) sample, its Tafel slope of 49 mV/dec is the lowest among the five SnTe samples. The exchange current densities ($j_0$) that reflect the inherent HER activity were determined by the extrapolated x-intercepts ($\eta=0$) of the fitted dash lines, as shown in Figure S3c (the inset shows higher magnification of a portion near the x-axis). The $j_0$ of the 70 nm SnTe (001) is the highest (1.065 µA cm$^{-2}$) among the five SnTe samples (the 20 nm, 49 nm, 94 nm, and 114 nm SnTe thin films have $j_0$ values of 0.534, 0.608, 0.559, 0.480 µA cm$^{-2}$, respectively) shown in Table 1. The corresponding log $j_0$ (A cm$^{-2}$) values[2] are also given in this table. The Nyquist plots of the five SnTe samples are given in Figure S3d. The charge transfer resistances ($R_{ct}$) of them are derived by applying a simplified Randles circuit model as shown in the inset of Figure S3d. The resulted $R_{ct}$ values are shown in Table 1 in which one can see that the 70 nm SnTe (001) exhibits the lowest $R_{ct}$ value of 19.84 Ω among the five SnTe samples, which is consistent with the observation that its $\eta$ and Tafel slope are the lowest and $j_0$ is the highest.

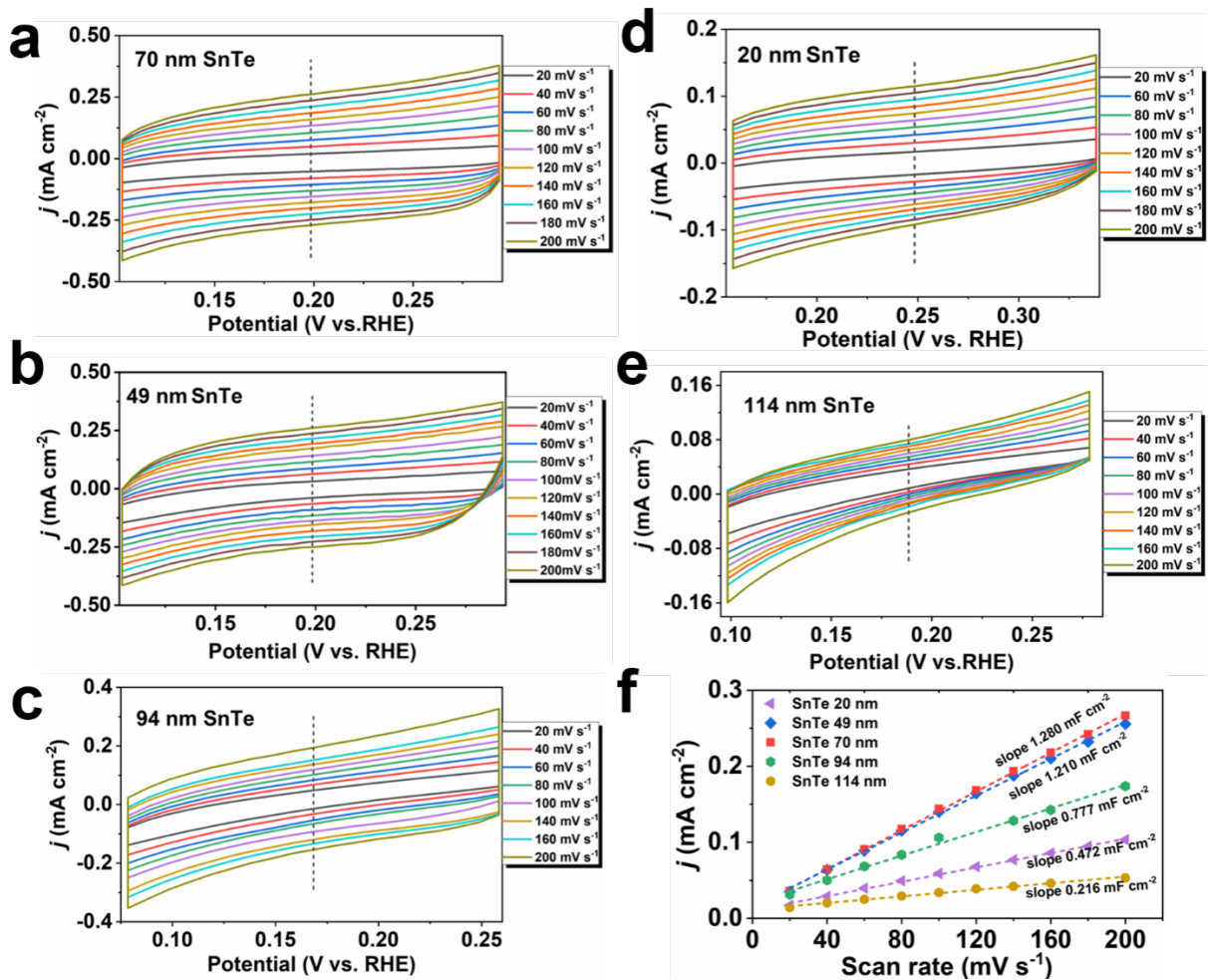

Figure S4 Electrochemical surface area measurements. Cyclic voltammograms (CV) curves with different scan rates of the (a) 70 nm SnTe, (b) 49 nm SnTe, (c) 94 nm SnTe, (d) 20 nm SnTe, (e) 114 nm SnTe, (f) Linear fits of the half capacitive currents as a function of scan rates for the extraction of $C_{dl}$ values of the five SnTe samples.

Remark of Figure S4:

Figure S4a-e shows the results of the CV measurements taken in the potential where no Faradic processes are observed, and these measurements aim to obtain the estimated electrochemical surface areas (ECSAs) of these samples. The capacitive current densities (($j_{anodic}$-$j_{cathodic}$)/2 taken at the potential values of the dash lines shown in Figure S4a-e) are plotted as a function of the scan rate as shown in Figure S4f. The resulting double-layer capacitances ($C_{dl}$) from the slopes of the lines shown in Figure S4f can be converted into the ECSA values using the specific capacitance value for a flat standard with 1 cm² of real surface area. The specific capacitance for a flat surface is generally found to be in the range of 20-60 µF cm$^{-2}$.[3-6] In the following calculations for obtaining the ECSA values, we take 40 µF cm$^{-2}$ as a moderate value, and define the $A_{ECSA}$ as:

$$A_{ECSA} = \frac{C_{dl}}{40 \text{ µF cm}^{-2} \text{per cm}^2_{ECSA}}$$

Thus, the ECSA values of the five SnTe samples can be calculated as shown below.

$$A_{ECSA}^{70 \text{ nm SnTe (001)}} = \frac{1.280 \text{ mF cm}^{-2}}{40 \text{ µF cm}^{-2}\text{per cm}^2_{ECSA}} = 32 \text{ cm}^2_{ECSA}$$

$$A_{ECSA}^{49 \text{ nm SnTe (001)}} = \frac{1.210 \text{ mF cm}^{-2}}{40 \text{ µF cm}^{-2}\text{per cm}^2_{ECSA}} = 30.25 \text{ cm}^2_{ECSA}$$

$$A_{ECSA}^{94 \text{ nm SnTe (001)}} = \frac{0.777 \text{ mF cm}^{-2}}{40 \text{ µF cm}^{-2}\text{per cm}^2_{ECSA}} = 19.425 \text{ cm}^2_{ECSA}$$

$$A_{ECSA}^{20 \text{ nm SnTe (001)}} = \frac{0.472 \text{ mF cm}^{-2}}{40 \text{ µF cm}^{-2}\text{per cm}^2_{ECSA}} = 11.8 \text{ cm}^2_{ECSA}$$

$$A_{ECSA}^{114 \text{ nm SnTe (001)}} = \frac{0.216 \text{ mF cm}^{-2}}{40 \text{ µF cm}^{-2}\text{per cm}^2_{ECSA}} = 5.4 \text{ cm}^2_{ECSA}$$

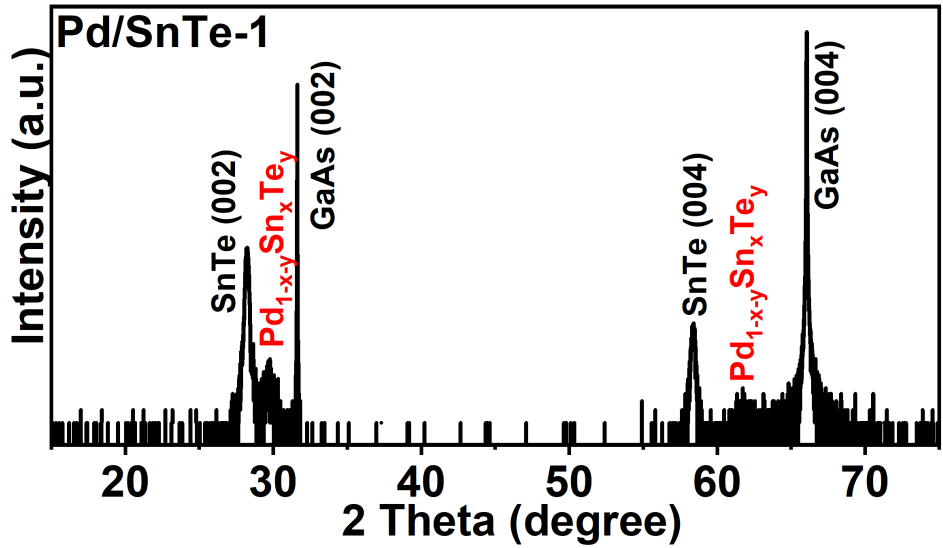

Figure S5. High-resolution X-ray diffraction 2θ-ω profile of the Pd/SnTe-1 sample.

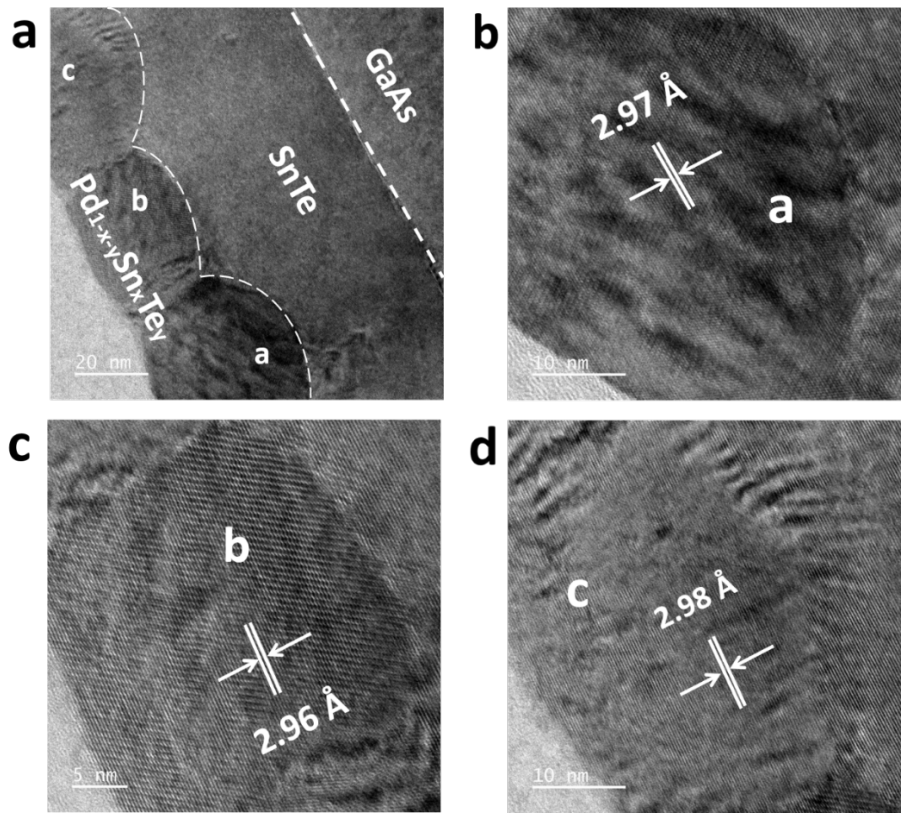

Figure S6. Cross-sectional high-resolution transmission electron microscopy (HRTEM) image of the Pd/SnTe-1 sample. (b)(c)(d) are the magnifying images of position a, b and c shown in (a).

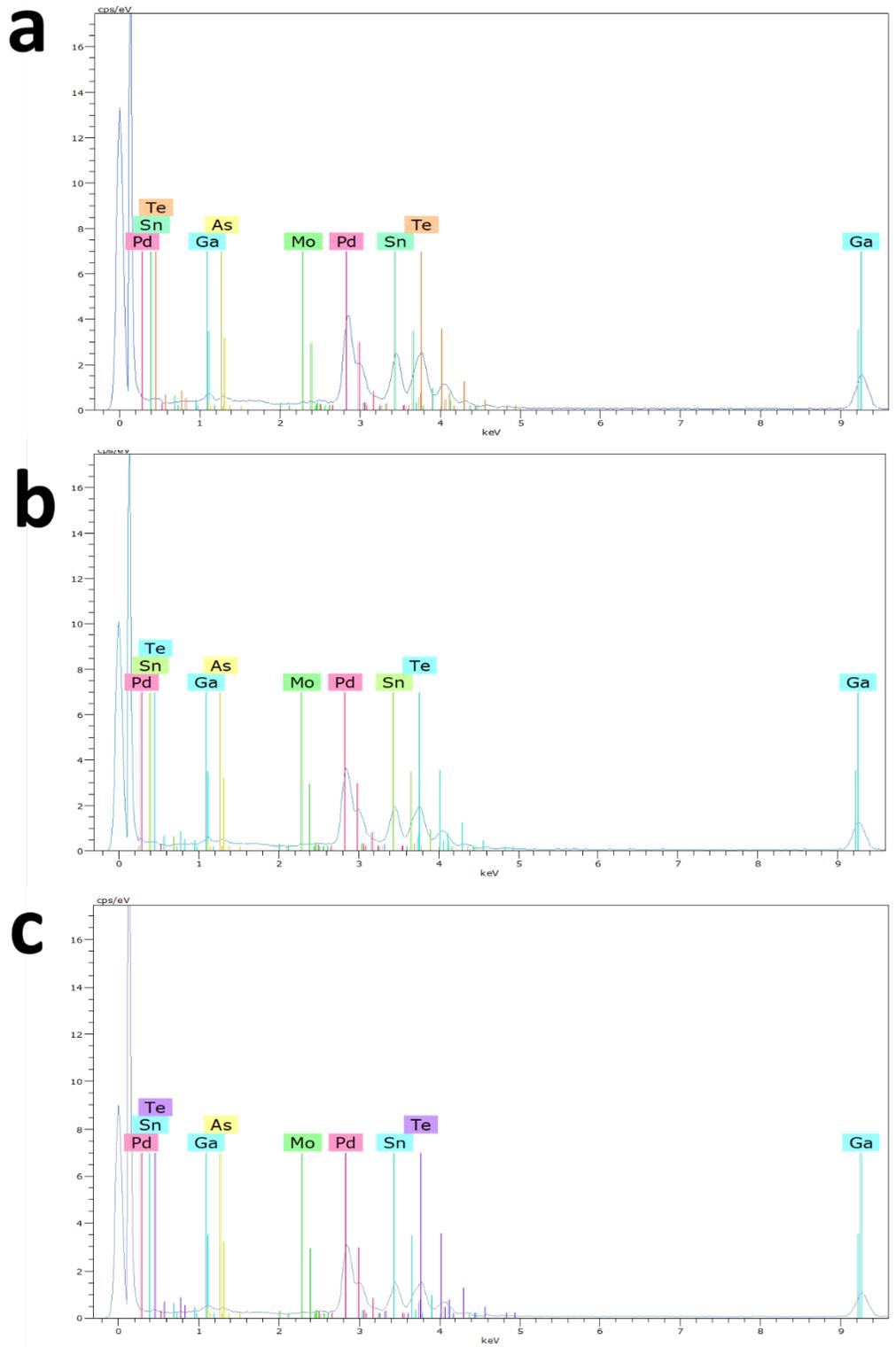

Figure S7. Energy dispersive spectroscopy (EDS) profiles of the position a, b and c shown in Figure S6, respectively.

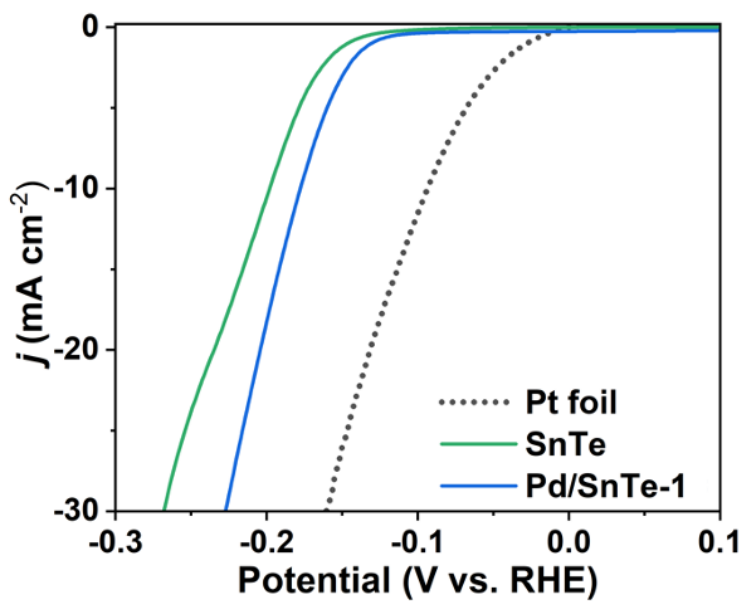

Figure S8. Polarization curves (*iR*-corrected) of the Pd/SnTe-1, SnTe (001) thin film, and a commercial Pt foil.

Supplementary note 1:

The $C_{dl}$ values can be converted into the ECSA using the specific capacitance value for a flat standard with 1 cm² of real surface area. Like what has been done earlier (as shown in Remark of Figure S4), in the following calculations for obtaining the ECSA values, we take 40 µF cm$^{-2}$ as a moderate value, and define $A_{ECSA}$ as:

$$A_{ECSA} = \frac{C_{dl}}{40 \text{ µF cm}^{-2} \text{per cm}^2_{ECSA}}$$

Thus, the ECSA values of the Pd(20nm)/SnTe and pure Pd(20nm) samples can be calculated as shown below.

$$A_{ECSA}^{20 \text{ nm Pd}} = \frac{0.320 \text{ mF cm}^{-2}}{40 \text{ μF cm}^{-2} \text{per cm}_{ECSA}^2} = 8.00 \text{ cm}_{ECSA}^2$$

$$A_{ECSA}^{20\text{Pd}/70\text{SnTe}} = \frac{0.505 \text{ mF cm}^{-2}}{40 \text{ μF cm}^{-2} \text{per cm}_{ECSA}^2} = 12.63 \text{ cm}_{ECSA}^2$$

Supplementary note 2:

To calculate the per-site TOF, we used the following formula:

$$\text{TOF} = \frac{\text{number of total hydrogen turnover/ cm}^2 \text{of geometric area}}{\text{number of active sites/ cm}^2 \text{ of geometric area}}$$

The number of total hydrogen turnover can be calculated according to (7):

$$\left(|j|\frac{\text{mA}}{\text{cm}^2}\right)\left(\frac{1\text{A}}{1000\text{mA}}\right)\left(\frac{1\frac{\text{C}}{\text{s}}}{1\text{A}}\right)\left(\frac{6.241\times10^{18}\text{e}^-}{1\text{C}}\right)\left(\frac{1\text{H}_2}{2\text{e}^-}\right) = \left(3.12 \times 10^{15} \frac{\frac{\text{H}_2}{\text{s}}}{\text{cm}^2}\right)|j|$$, where $j$ is taken from the polarization curves.

In general, the HER activity of a sample is the average activity from the various sites of the sample and Pd atoms should be the active sites of both the pure Pd(20nm) sample and the Pd(20nm)/SnTe heterostructure. For samples with a rough surface, the number of active sites per unit surface is usually estimated to be the 2/3 power of the ratio of the number of Pd atoms in a Pd unit cell over the unit volume.[8–10]

The number of the active sites of both the pure Pd sample and the Pd layer on the heterostructure per real surfaces area can be calculated using the parameters presented below:

| | |
|---|---|
| Pd unit cell | 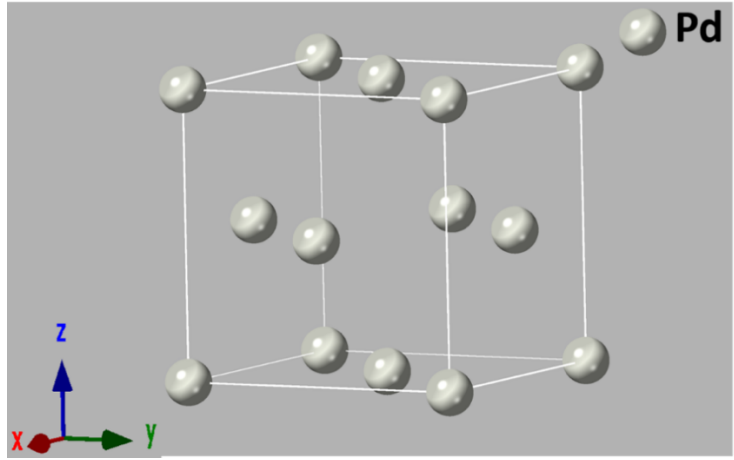 |
| Volume | 58.855 Å$^3$ / unit cell |
| Contains | 4 Pd atoms / unit cell |

$$\text{Active sites}_{Pd} = \left(\frac{4\ \text{atoms/unit cell}}{58.855\ \text{Å}^3/\text{unit cell}}\right)^{\frac{2}{3}} = 1.6654 \times 10^{15}\ \text{atom cm}^{-2}_{\text{real}}$$

In reference to what has been described in Supplementary note 2, the TOF values of the pure Pd sample and the Pd/SnTe heterostructure can be calculated by the following equation:

$$\text{TOF} = \frac{3.12 \times 10^{15} \frac{H_2/s}{cm^2}}{1.6654 \times 10^{15}\ \text{atom cm}^{-2}_{\text{real}} \times A^{Pd}_{ECSA}} |j|.$$

Supplementary note 3:

To further support our argument, we calculated the free energies of hydrogen adsorption ($\Delta G_H$) on the Pd/SnTe heterostructure whose hydrogen coverages are 1/9 ML, 2/9 ML, and 3/9 ML. Figure S9(a) shows the most active adsorption site and Figure S9(b,c) show the less active but more stable adsorption sites. Figures S9 (d) and (e) show the hydrogen coverage at 2/9 ML and 3/9 ML, respectively. In Figures S9 (d, e), after hydrogen atoms occupy the more stable adsorption sites, they are adsorbed at the less stable but more active adsorption sites. $\Delta G_H$ of Figures S9 (d, e) is nearly the same as that of Figure S9 (a), indicating that the less stable but more active adsorption sites determine the HER performance. It is worth to note that as shown in Figure S9(a,d,e), with increasing hydrogen coverage from 1/9 ML to 3/9 ML, $\Delta G_H$ at the most active adsorption site does not change much, indicating that the different hydrogen coverages do not significantly affect the HER performance.

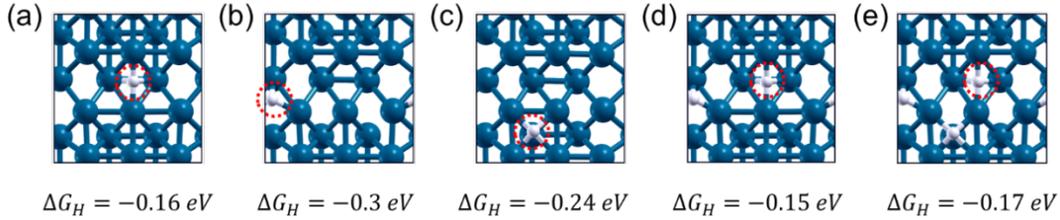

(a) $\Delta G_H = -0.16\ eV$  (b) $\Delta G_H = -0.3\ eV$  (c) $\Delta G_H = -0.24\ eV$  (d) $\Delta G_H = -0.15\ eV$  (e) $\Delta G_H = -0.17\ eV$

Figure S9. Top view of adsorption structures and free energies of hydrogen adsorption ($\Delta G_H$) on the Pd/SnTe heterostructure. The hydrogen coverages are (a-c) 1/9 ML, (d) 2/9 ML, (e) 3/9 ML. $\Delta G_H$ is calculated at the hydrogen adsorption sites labeled by the red dashed circles. (a), (d), and (e) show the most active adsorption sites, while (b) and (c) show the less active but more stable adsorption sites.

Supplementary note 4: formation energy calculation

Our experimental results suggest that Pd overlayers grown on the partially oxidized SnTe (001) layer had better crystalline morphology than those on the pure SnTe (001) layer, and no $Pd_{1-x-y}Sn_xTe_y$ was formed. Here, we calculated the formation energies of Pd (100) overlayers grown on pure and oxidized SnTe layer using DFT.[11]

For a pure SnTe layer:

$$E_f = (E_{nPd/SnTe} - E_{SnTe} - 3/2 \cdot E_{Pd})/3$$

where $E_{nPd/SnTe}$ is the total energy of the Pd/SnTe heterostructure with 3 ML Pd and 6 ML SnTe (001), $E_{SnTe}$ is the energy of the relaxed 6 ML SnTe (001) slab, and $E_{Pd}$ is the energy of 2 ML Pd (100) in the bulk phase.

For an oxidized SnTe layer:

$$E_f = (E_{nPd/SnTe_{oxi}} - E_{SnTe_{oxi}} - 3/2 \cdot E_{Pd})/3$$

Our simulations suggest that Pd tends to bond with Sn and cannot retain the Pd (100) structure if Pd is directly deposited on a pure SnTe layer. When SnTe is oxidized, the formed $SnO_2$ may prevent Pd from bonding with Sn, which lowers $E_f$ and thus promotes the formation of Pd overlayers.

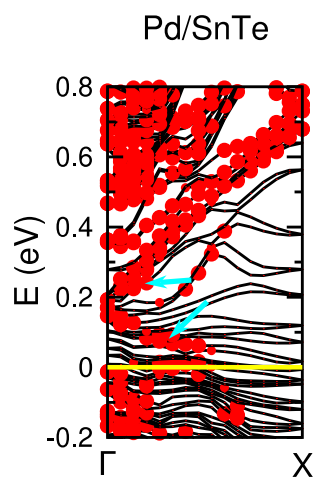

Figure S10. Band structure of Pd/SnTe heterostructure. Red dots indicate the contributions from the second layer of SnTe (001) upper surface covered by Pd overlayers. The TSSs labelled by cyan arrows can be seen in all the band structures shown here. The Fermi level is set to zero.

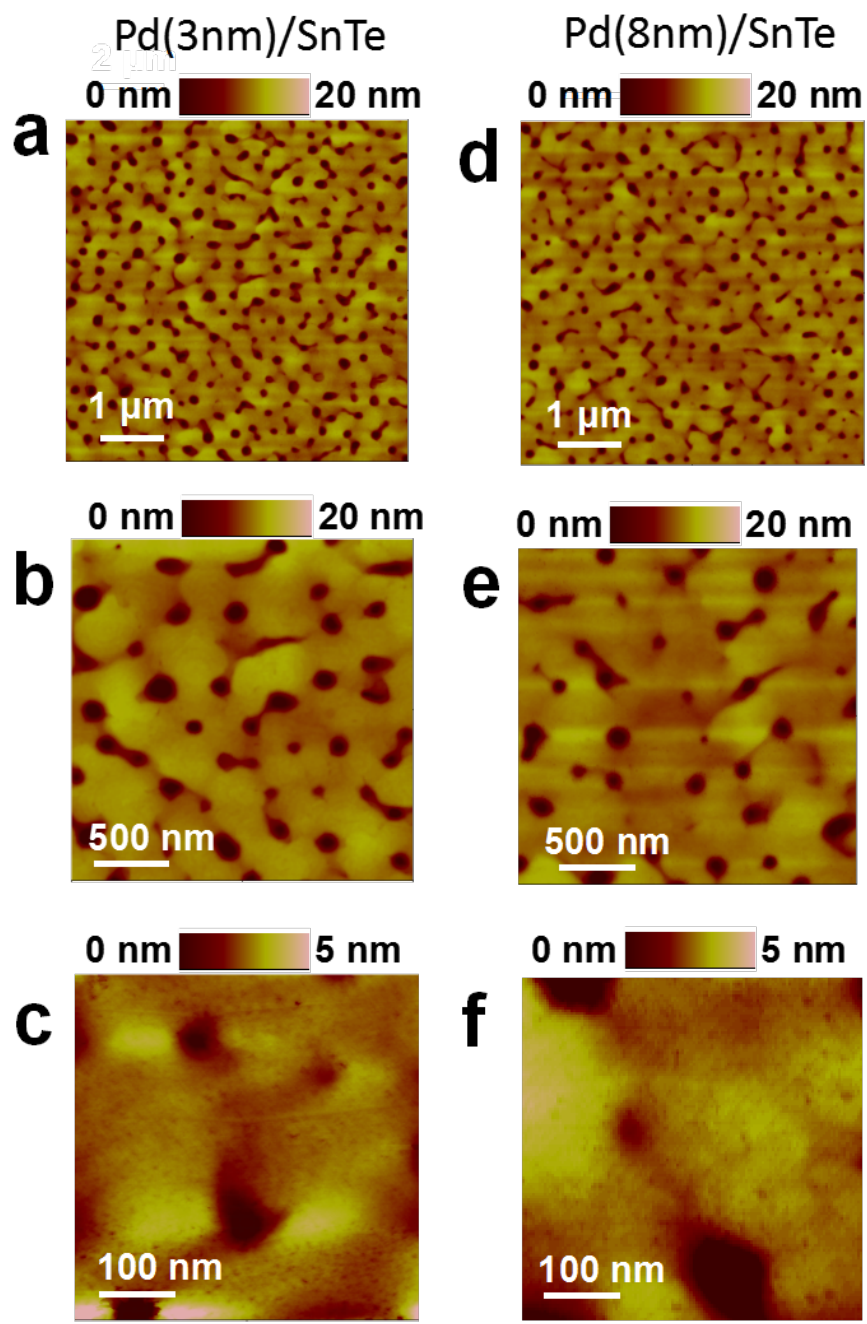

Figure S11. AFM images with an increasing magnification of (a-c) Pd(3nm)/SnTe and (d-e) Pd(8nm)/SnTe heterostructures. The surfaces with some pores indicates the Pd layers should be discontinuous.

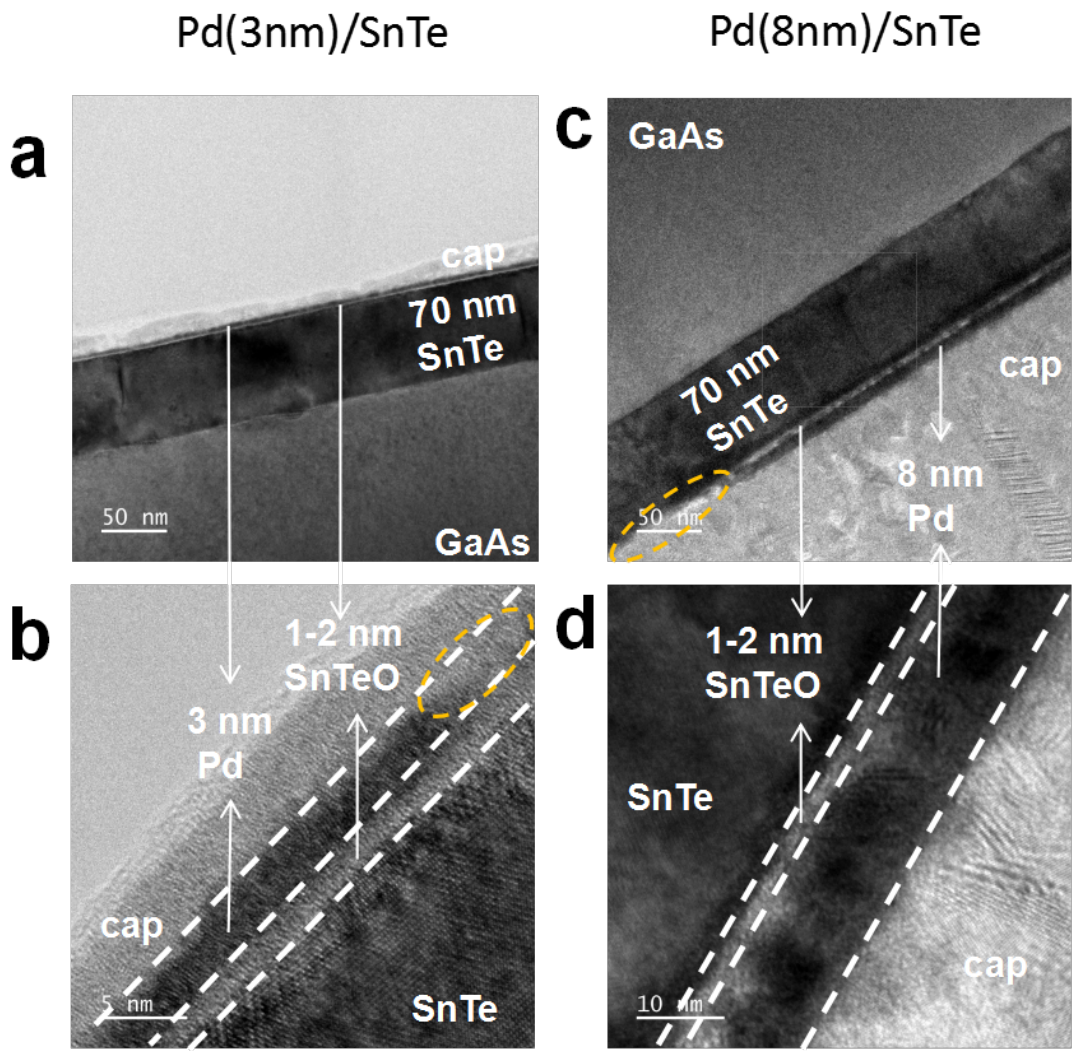

Figure S12. Cross-sectional HRTEM images with an increasing magnification of (a-b) Pd(3nm)/SnTe and (c-d) Pd(8nm)/SnTe heterostructures. The yellow circles indicate the discontinuous nature of Pd layers of the two heterostructures.